\documentstyle[aps,preprint]{revtex}
\begin{document} \title{Static magnetic response of clusters in
$Co_{0.2}Zn_{0.8}Fe_{1.95}Ho_{0.05}O_{4}$ spinel
oxide}
\author{R.N.Bhowmik\footnote{e-mail:rnb@cmp.saha.ernet.in} and
R.Ranganathan} \address{Saha Institute of Nuclear Physics, Experimental
Condensed Matter Physics Division\\
1/AF, Bidhannagar, Calcutta-700064,
India\\}
\author{R. Nagarajan} \address{Tata Institute of Fundamental
Research, Department of Condensed Matter Physics and Materials Science,
Colaba
Road, Mumbai, India\\}
\maketitle
\begin{abstract}
Earlier investigations of Co$_{0.2}$Zn$_{0.8}$Fe$_2$O$_4$ spinel oxide has
shown the existence of ''super-ferromagnetic'' clusters containing Fe$^{3+}$ 
and Ho$^{3+}$ ions along with small size clusters of Fe$^{3+}$ ions 
(Bhowmik {\it et al.}, J. Magn. Magn. Mater. \textbf{247}, 83 (2002)). Here, 
we report the static magnetic response of these clusters. The experimental
data suggest some interesting magnetic features, such as, enhancement of
magnetization; re-entrant magnetic transitions with paramagnetic to
ferromagnetic state below T$_C$ $\approx$ 225 K and ferromagnetic to spin
glass like state below T$_m$ $\approx$ 120 K; appearance of field induced
ferromagnetic state. We also observe an unusual maximum in the thermoremanent 
magnetization (TRM) vs temperature data. Our measurements suggest that this 
unusuality in TRM is related to the blocking of ''super-ferromagnetic'' 
clusters, out of the ferromagnetic state, along their local anisotropy axis.
\end{abstract}

\section{introduction} In present scenario of
condensed matter physics rare earth ions are playing an active role in
magnetic oxides and this is intensively 
studied in case of perovskites and
pyrochlores \cite{Teresa,Alanso}. The spinel oxides with formula unit
AB$_2$O$_4$ \cite{Blasse} represent another most important and interesting
class of magnetic materials. The magnetic
disorder and exchange frustration in spinel structure, introduced by size 
mismatch of cations and competition between superexchange 
interactions amongst A and B site moments, gives rise to various kinds of 
magnetic order \cite{Bel}. Inspite of the enourmous substitution works in
spinel oxide \cite{Blasse,Dorman}, less attention has been paid for the 
substitution of rare earth (RE) ions in spinel oxide. However, the theoretical 
and experimental investigation of RE substitution are also equally important, 
as given for transition 
ion substitution in spinel, for the understand of coupling effect between 3d-4f 
spins in spinel oxide. For example, the vector mean-field model 
(Heisenberg) \cite{Gabay} predicts the appearance of a re-entrant phase in
a disorder magnet \cite{Bel,Dho} where the ferromagnetic order of longitudinal 
spin components coexists with the spin glass order of transverse spin 
components. On the otherhand, strong crystalline electric field effect of rare 
earth (RE) ions rotate the spins along its local anisotropy axis and
therefore, gives an Ising character to the spins. Consequently, the critical 
phenomena like magnetic transitions with AT and GT lines in rare earth
containing disorder magnetic systems are expected to be different in 
comparison with the isotropic (Heisenberg) spin glass behaviour of transition 
metal ions \cite{Mydosh}. Therefore, a systematic and detailed investigations 
of rare earth substitution are very essential over a wide range of spinel 
oxides, with different magnetic structure such as long range 
ferro/ferrimagnet, antiferromagnet, spin glass/cluster spin glass etc.\\
In most of the earlier known cases, rare earth (RE) ions were
substituted in a long range ferrimagnetic spinel oxides like Ni-Zn
ferrite, Co-ferrites \cite{Rezlu,Cheng}. The lack of sufficient
experimental reports dealing with rare earth substitution in a frustrated and
magnetically diluted spinel oxide motivated us to study the effects of rare 
earth ions in Co$_{0.2}$Zn$_{0.8}$Fe$_2$O$_4$ spinel oxide. This system
with cations distribution 
(Zn$^{2+}_{0.8}$Fe$^{3+}_{0.2}$)$_A$[Co$^{2+}_{0.2}$Fe$^{3+}_{1.8}$]$_B$O$_4$
(A: tetrahedral site, B: octahedral site] is highly A site
magnetically diluted, whereas the B site moments form finite clusters
and the spins inside the clusters form canted structure \cite{rnbzn8}. If
RE$^{3+}$ ions are introduced in this system, the B site Fe$^{3+}$ ions
will be replaced due to strong B site occupancy of RE$^{3+}$ ions and we assume
that some of the clusters will contain the RE$^{3+}$ ions. Aside from the
large values of free ion magnetic
moment ($\sim$ 10 $\mu_{B}$ for Ho$^{3+}$, Dy$^{3+}$
etc.), the competition between the single ion anisotropy of RE$^{3+}$ ions
and random field anisotropy of the spin canting states may strongly modify the
cluster glass properties of Co$_{0.2}$Zn$_{0.8}$Fe$_2$O$_4$ \cite{rnbzn8}.\\
With the above expectation, we prepared polycrystalline
samples of Co$_{0.2}$Zn$_{0.8}$Fe$_{2-x}$RE$_x$O$_4$ ({\it x} $\sim$ 0.05;
RE = Dy, Ho and Er) using conventional solid state method
and compared their magnetic properties \cite{rnbjac}.
It was observed that the magnetic properties of these samples are almost
similar irrespective of Dy, Ho and Er ions substitution in place of
Fe$^{3+}$ ions. A detailed AC susceptibility measurements \cite{rnbHo5jm} on
Co$_{0.2}$Zn$_{0.8}$Fe$_{1.95}$Ho$_{0.05}$O$_4$ spinel oxide have indicated the
following interesting magnetic behaviours \cite{rnbHo5jm}: (i) A re-entrant 
like magnetic behaviour with paramagnetic$\leftrightarrow$ ferromagneic
$\leftrightarrow$ cluster spin glass order, (ii) Existence of two types of
clusters, {\it viz}, "super-ferromagnetic cluster" and small clusters
associated with two potential barriers as 
E$_{\pm}$ = $\frac{E_{B}}{H^{2}}$[H$\mp$ 2K/M$_S$]$^2$ \cite{rnbHo5jm}, and
(iii) Unusual TRM maximum at $\approx$ 170 K comparing the conventional
magnetic materials. In this communication we present
the static magnetic response of the clusters, {\it i.e.}, small
clusters and "super-ferromagnetic" clusters in
Co$_{0.2}$Zn$_{0.8}$Fe$_{1.95}$Ho$_{0.05}$O$_4$ spinel oxide, by
performing systematic dc magnetization measurements as a function of 
temperature, magnetic field and time.

\section{EXPERIMENTAL}
We have carried out the low field ($<$ 100 Oe) dc magnetization measurements 
using home made magnetometer \cite{Anindita} and high field ($\geq$ 100 Oe) 
magnetic measurements using SQUID (Quantum Design MPMS) magnetometer and 
vibrating sample magnetometer (Oxford Inc.). The dc magnetization was measured 
from 10 K to 300 K under zero field cooled (ZFC) and field cooled (FC) modes.
In ZFC (FC) mode the sample was cooled in absence (presence) of magnetic field 
from 300 K to the measurement temperature (e.g. 10 K), then constant dc field 
(say, 30 Oe) was applied and magnetization data were recorded while increasing
temperature. In FC mode the cooling field and measurement fields are same in
magnitude. The field dependence of magnetization was studied under ZFC mode.
The experimental procedure of the time dependence of magnetization (relaxation 
experiment in ZFC and FC mode) and the temperature dependence of 
thermoremanent magnetization (TRM) are described in discussion section.

\section{RESULTS AND DISCUSSION}
\subsection{Temperature dependence of magnetization}
Fig. 1 shows the comparative ZFC magnetization data at 30 Oe for
Co$_{0.2}$Zn$_{0.8}$Fe$_{2-x}$Ho$_x$O$_4$ (x = 0, 0.05, 1.0) system.
The immediate striking feature is that introduction of 0.05 concentration (x) 
of Ho enhances the magnetization by a factor of about 7 compared to that of
undoped sample. We believe that this enhacement primarily arises due to large 
moment of Ho$^{3+}$ (free ion moment $\approx$ 10 $\mu_{B}$) which is 
replacing the mement of Fe$^{3+}$ (free ion moment $\approx$ 5 $\mu_{B}$). 
Larger substitution of Ho, instead of further enhancement of magnetization, 
results in a reduction of magnetization, though it is still higher than
that for x = 0 sample. This observations are consistent with our argument
\cite{rnbjac} that at higher concentration (x), all the Ho ions are not 
incorporated in the lattice but results in an impurity phase with low magnetic
moment. We do see that substitution of even 0.05 concentration of Ho, 
suppresses the short range ferrimagnetic transition at T$_{m2}$ $\approx$
260 K occuring in the undoped sample.\\
The low field dc susceptibility (M/H) in Fig. 2 not only shows
irreversibility below 220 K between ZFC (Fig. 2a) and FC (Fig. 2b) 
magnetizations, the temperature dependences are distinctly different (shown
together in Fig. 2a inset for 0.2 Oe). While the temperature dependence of ZFC
magnetization shows a peak (Fig. 2a), 
the FC magnetization shows a distinct plateau region (Fig. 2b)
which is typical of a re-entrant spin glass system \cite{Dho}. We note that the
temperatures at which plateau region begins and ends ($\sim$ 100 K to $\sim$ 
175 K) at low fields nearly match the temperature of the two peaks T$_{L}$
$\approx$ 90 K and T$_{H}$ $\approx$ 150 K seen in 
our $\chi^{\prime\prime}$ of AC susceptibility \cite{rnbHo5jm}.  
At present the reason for an increase in both ZFC and FC magnetizations  
below 50 K (Fig. 2), which is more prominent in FC magnetization, 
is not clear to us. 
However, the appearance of magnetization minimum at T$_{min}$ can be
attributed to different type of compensation effects, {\it e.g.}, the
compensation of antiferromagnetic interactions by ferromagnetic
interactions \cite{Dho,Gal}; the compensation of magnetization and anisotropy 
constants arising in multi cations 
(Ho$^{3+}$, Fe$^{3+}$ and Co$^{2+}$) system \cite{Okoshi}.\\  
At fields H $\geq$ 10 Oe, the following features: (i) magnetic irreversibility 
between FC and ZFC magnetization below a temperature T$_{irr}$ with a 
predominant magnetization maximum about T$_m$, and (ii) field dependence of 
both T$_{irr}$ and T$_{m}$ suggest spin glass behaviour \cite{Mydosh} in our 
system. For spin glass system, T$_m$ and T$_{irr}$ $\propto$
H$^n$ \cite{Mydosh}. Our data for H $\leq$ 100 Oe, fits to a single value of 
n = 0.46 (Fig. 3 inset). For H $\geq$ 500 Oe, T$_{irr}$ (H) deviates 
drastically from this exponent value, which may be
due to occurrence of field induced ferromagnetism. 
Interestingly, the value 0.46 of the exponent is closer to the vlaue 
0.58 found for La$_{0.5}$Sr$_{0.5}$CoO$_3$ where cluster spin glass is
suggested to coexist with ferromagnetic order \cite{Nam},
rather than those predicted by infinite range vector mean-field model n = 0.67 
(for De Almeida-Thouless (AT) line) and n = 1 (for Gabay-Touless (GT) line)
\cite{Gwyn}. We have also analysed the T$_m$ (H) data
and find the exponet n = 0.023 $\pm$ 0.002 (H $\leq$ 100 Oe).
This value is even less than that found for T$_{irr}$ (H). Therefore, our
system belongs neither to typical Heisenberg class (exhibits both AT and GT 
line) nor typical Ising class (exhibit only AT line). However,
a randomly anisotropic Heisenberg SG has also shown to exhibit only AT line 
only at low field range \cite{Gwyn}. Our fit value of n (for T$_{irr}$ (H)) is
consistent with n = 0.48$\pm$ 0.01 seen for a short-range, including
anisotropy effect, 3D Ising SG system \cite{Fotis}. Hence, we 
attribute the deviation of our n values from those of infinite range 
vector mean-field predicted AT line or GT line, primarily, to the effect of 
field induced ferromagnetic order in its coexistence with spin
glass state. The appearance of only AT type behaviour of T$_{irr}$ (H), we 
believe, due to the anisotropy effect of Ho$^{3+}$ ions. 
The 
field induced effect on the magnetization minimum at T$_{min}$ (Fig. 3) can be 
quantified by taking the difference
$\Delta$MFC of FC mangnetization at T$_m$ (MFC$_{max}$) and that at
T$_{min}$ (MFC$_{min}$), normalized to MFC$_{max}$, {\it i.e.}, 
$\Delta$MFC = (MFC$_{max}$-MFC$_{min}$)/MFC$_{max}$. We find that $\Delta$MFC
decreses from 60\% at 1 Oe to 1.2 \% at 1 Tesla. 
This clearly shows that the field induced ferromagetic order, which dominates 
over the spin glass states in this system, also takes into account for 
increasing flatness of M vs T curve with increasing field \cite{Hesse}.\\ 
Because of the field induced FM order, it is not easy to determine the magnetic 
ordering temperture (T$_C$) in a magnetically disordered system such as ours. 
From the first order derivative of the real part of AC suscpetibility vs T 
curve, we derived possible value of T$_C$ as $\sim$190 K \cite{rnbjac}. 
However, the dc magnetization data (Fig. 2a inset) show
magnetic irreversibility even upto 220K, suggesting T$_C$ is $\geq$ 220 K. 
We have attempted to estimate T$_C$ from Curie-Weiss law $\chi_{dc}$ =
C/(T-$\theta_{p}$), from the dc susceptibility ($\chi_{dc}$) data at 100 Oe. We 
find that the data above 275 K gives a good fit (inset of Fig. 3) for 
Curie-Weiss law with Curie constant C $\approx$ 0.14 and paramagnetic curie 
temperature $\theta_{p}$ $\approx$ 240K.
The positive value of $\theta_{p}$ confirms the dominant ferromagnetic 
interactions in our system.

\subsection{Field dependence of magnetization}
Fig. 4 shows the magnetic hysteresis of the sample over the
field range -1.5 Tesla to +1.5 Tesla. The wide hysteresis loop at 10
K with a coercive field (Hc) $\sim$ 0.25 Tesla and a
remenant magnetization (M$_R$) $\sim$ 32 emu/g 
confirm the presence of ferromagnetic interactions in this
material at low temperature. The hysteresis loop area
drastically reduces on increase of temperature with H$_C$ and
M$_R$ becoming small above 50 K and no hysteresis loop is observed at T $\geq$
250K. The sharp increase of H$_C$ below 100K also
takes into account the effect of coexisting spin glass state
at low temperature. The small peak in M$_R$ around 170 K and small increase in
H$_C$ around 160K are most probably related to anisotropy contribution of
Ho$^{3+}$ ions \cite{Gesev}.\\ 
Fig. 5 shows magnetization (M) vs field (H) data at selected
temperatures with H upto 12 Tesla. In the temperature range 10-160K, the 
M shows an initial rapid increase, followed by slower increase 
with H which is a typical ferro or ferrimagnetic response of the sample. 
The lack of saturation of M upto 12 Tesla suggest that the 
sample is not an infinite long range order ferromagnet \cite{Gal,Mira}. The 
isotherms can represent either a system with canted spin structure
in the ferromagnetic cluster \cite{Nam} or a system where ferromagnetic order
coexists with superparamagnetic component \cite{Gesev}.
In order to determine the spontaneous magnetization (M$_S$) due
to ferromagnetism, we applied Arrot plot (M$^2$ vs H/M) (Fig. 5b). 
Although M(H) at 250 K and 305 K show non-linear increase, 
Arrot plot gives no M$_S$ for these 
temperatures, which indicate that at T $\geq$ 250 K, the system is
in a paramagnetic regime mixed with short range interacting clusters.
We fit M$_S$(T) data to the functional form: 
M$_S$(T) = K(T$_C$-T)$^{1/\gamma}$ for T $< T_C$, as has been applied for
disorder ferromagnet \cite{Bel}. 
The best fit was obtained with K = 3.646, $\gamma$ = 0.55$\pm$0.01 and Curie
temperature T$_C$ = 225$\pm$5 K. The inset of Fig. 5b also shows that the
obtained value of M$_S$ deviates and remain lower below 100 K than the fitted 
curve using above equation. This indicates the canted spin structure in those
clusters which show low temperature spin glass freezing.\\ 
To verify the originality of the observed peak temperatures in AC 
susceptibility and dc magnetization data, the non-linear
response of the sample is examined at low field
(H $\leq$ 100 Oe). The dc susceptibility
($\chi_{dc}$= M/H) consists of linear (independent of field) and non-linear 
(dependent on applied field) components as
\begin{equation}
\chi_{dc} = \chi_{1} - \chi_{3}H^{2} + \chi_{5}H^{4} - .
\end{equation}
The linear component ($\chi_{1}$) was obtained from $\chi_{dc}$ vs H plot
at the zero field limit. $\chi_{3}$ and $\chi_{5}$ are the
non-linear components of dc susceptibility.
The non-linear susceptibility is defined as $\chi_{nl}$ =
$\chi_{1}$-$\chi_{dc}$.
The non-linear components were obtained
by polynomial fit of $\chi_{nl}$/H$^{2}$ (y axis) vs H$^{2}$ (x axis)
plot at each temperature.
For discussion, we neglect $\chi_{5}$ and higher non-linear components as 
their magnitudes are negligibly small.
Fig. 6a shows a broad maximum in $\chi_{1}$ at $\approx$ 110 K and clear
peak at $\approx$ 166 K. Eventhough, there are the
signatures of two peaks at $\approx$ 200 K and 227 K, respectively, but they
are of the order of the limit 10\% error.
However, the temperature dependence of $\chi_{3}$ (Fig. 6b) shows four
clear peaks at $\approx$ 105 K, 166 K, 200 K, 230 K, respectively.  
The low temperature peak at $\approx$ 105 K may be associated with the cluster 
spin freezing below $\approx$ 120 K \cite{rnbHo5jm}.
Some specific features of $\chi_{3}$ at 105 K, {\it i.e.}, more divergence than
other three peaks and rapid decrease of $\chi_{1}$ below 105 K suggest 
low temperature re-entrant magnetic phase, as predicted in Ising model
\cite{Gwyn}. The fourth one at $\approx$ 230 K is close to the Curie 
temperature (T$_C$ $\approx$ 225$\pm$5 K) obtained from Arrot plot (Fig. 5b)
and, so associated with the paramagnetic to ferromagnetic phase transition of
the sample.
The $\chi_{3}$ peak at $\approx$ 166 K is close to the high temperature
$\chi^{\prime\prime}$ peak at T$_H$ $\approx$ 160 K \cite{rnbHo5jm} and
we suggest its origin from the domain rotation effects \cite{Gwyn} of
"super-ferromagnetic" clusters.
Finally, the $\chi_{3}$ peak at $\approx$ 200 K is near to the
value of T$_C$ $\sim$190 K obtained from the first order derivative of
$\chi^\prime$ vs T data \cite{rnbjac} and its origin is not clear at present. 

\subsection{Time dependence of magnetization}
To study the time response of the magnetic clusters at different temperatures
below 250 K, we cooled the sample from 300 K under ZFC and FC mode. In view of
the unusual increase of TRM above 150 K and with peak at $\sim$ 170 K (see
next section, Fig. 8) in this sample, we have investigated time response of
magnetization at 153 K (ZFC) and 160 K (FC) to see if the increase in TRM has
any effect on the time response also. 
160 K is chosen with the expection that it would not make much difference with 
respect to 153K, and if at all it would show some effect, a slightly higher 
temperature is taken to ensure the effect, if any, of the TRM maximum. 
The sample was waited for 300 sec at
153 K (ZFC) and at 160 K (60 Oe FC) prior to application and removal of 60 
Oe field, respectively. Fig. 7a shows that the magnetization at 153 K 
in the presence of field achieves equlibrium quite fast and remains
constant during the observation time upto 5$\times$10$^4$ sec.
On the other hand, the remanent magnetization (M$_R$) 
at 160 K decreases with time (Fig. 7a) following power law behavior 
\begin{equation}
M_R = M_0t^{-\alpha}
\end{equation}
where t is the time of observation and $\alpha$ is the exponent.
Thus, in the time dependence, we did not observe any effect
of maximum in TRM vs T. Before confirming about no effect of maximum 
in TRM, we have applied almost identical condition to the sample for
M$_R$ vs t measurement at 160 K which was given during TRM vs T measurement.
In this process, the sample was field cooled (60 Oe) from 300 K to 38 K. Then
field was removed and M$_R$(t) was recorded for 6$\times$10$^3$ sec. After
measurement, sample was slowly warmed up to 160 K in the absence of external 
field and the M$_R$(t) was observed for 4.5$\times$10$^4$ sec. Eventhough the
condition is identical, M$_R$ does not increase with time 
(Fig. 7a), as observed with temperature (Fig. 8). Fig. 7a also shows
that the M$_R$ at 160 K, after thermal cycling, is lower in comparison with
earlier data at 160 K (Fig. 7a). Interestingly, M$_R$ still follows power
law, but with larger value of exponent. Therefore, inspite of the fast 
equilibrium at 153 K (ZFC), the magnetic relaxation at 160 K suggests that a
fraction of clusters, out of ferromagnetic state, are blocked in this
temperature regime and showing relaxation effect.
The time dependence of M$_R$, after direct 60 Oe field cooling, for all other 
temperatures in the range 20 K to 250 K (data not shown) also follow power law
behaviour. The values of fit parameters are shown in Fig. 7b. The decrease of 
M$_0$ with increase of temperature is expected due to thermally
activated processes. The application of power law decay and almost linear 
decrease of exponent $\alpha$ below 100 K indicates that spin glass 
state of the sample is of Ising type \cite{Kin}. The decrease of the exponent
below 100 K implies that magnetic relaxation becomes very slow due to freezing
of clusters, whereas the decrease of the exponent above 100 K is well
explained by the blocking of a fraction of
clusters in the ferromagnetic state. The ferromagnetic clusters quickly
reach to its equilibrium magnetization because of very small coercive field 
(inset of Fig. 4b), where as the clusters in the blocking state slowly relax.
Above 250 K the system is in the paramagnetic state and all magnetic clusters 
instantly reach its equilibrium value (M$_0$) and hence $\alpha$ again tends 
to zero value.
\subsection{Thermoremanent magnetization}
Thermoremanent magnetization (TRM) vs temperature (T) has been measured by 
cooling the sample from 300 K to 10 K in presence of constant magnetic field 
and after reducing the cooling field to zero value.
The unusual observation of a maximum in the TRM(T) data at T$^{max}_{trm}$ 
$\approx$ 170K (for 30 Oe cooling field) \cite{rnbHo5jm} has generated further 
interest for the present sample. To observe the effect of higher applied 
fields ($\sim$ 500 Oe, 1 kOe and 1 Tesla) on TRM vs T, SQUID magnetometer was
employed. Care was taken to ensure that the remanent field of superconducting 
magnet is as close to zero as possible, by measuring the remanent field with a
known paramagnetic sample and then applying a compensating field during TRM
measurement. The TRM data (normalized by colling field: H$_{FC}$) (Fig. 8) 
show a maximum about 180 K and a minimum
about 125 K for H = 500 Oe. It is also observed that the maximum in TRM is 
suppressed with increasing cooling field to 1 Tesla. Consequently, the 
minimum in TRM about 125 K transform into a plateau in the temperature
range 100 K to 175, which is a character of re-entrant magnetic behaviour
\cite{Murali}. It is also found for 500 Oe that the increase of TRM at 180 K
with respect to TRM at 125 K is only 0.4\% of the decrease of
TRM at 10 K with respect to TRM (125 K). This confirms that 
a spin glass state, coexisting with ferromagnetic state, at low temperature 
sharply decreases above
125 K and a superparamagnetic state of a fraction of clusters 
(super-ferromagnetic) develops in the
ferromagnetic state above 125 K which we postulated in this sample
\cite{rnbHo5jm}. Eventhough the contribution of
superparamagnetic type clusters is about 0.4 \%, but their existence in
the sample is intrinsic. We suggest that the minimum in TRM about 125 K arises 
due to the competition effects between spin glass component and 
super-ferromagnetic component of the clusters. 
The "super-ferromagnetic" clusters, even, show spontaneous magnetization in
absence of external field due to their strong internal field 
(H$_i$ = 2K/M$_S$) and below 180 K these clusters are more relaxed (blocked) 
in different local anisotropy axes related to Ho$^{3+}$ ions. 
In Ref. \cite{rnbHo5jm}, we demonstrated the freezing of 
"super-ferromagnetic" clusters due to the antiferromagnetic inter-cluster
interactions. The larger negative TRM with assymetry behaviour for 
H$_{FC}$ = -1 kOe with respect to that taken for +1 kOe (inset of Fig. 8)
indicates that really, blocking of the 'super-ferromagnetic" clusters along its
local anisotropy axes prefers antiferromagnetic direction with respect to the
ferromagnetic order of large number of small clusters.
We, further, suggest that the magnetic behaviour of the "super-ferromagnetic" 
clusters will be governed by Ho$^{3+}$ mement due to its large magnetic moment 
and anisotropy. If we consider the "super-ferromagnetic" cluster as single
domain particle, the effective spin of this cluster may add Ising nature to
the sample \cite{Kin} which is reflected in time dependence of field cooled
remanent magnetization and in T$_{irr}$ (H) fit. At this stage, we suggest
that the re-orientation effect of Ho$^{3+}$ spins \cite{Gal,Gesev} below 125 K
are the cause of ferromagnetic increase of magnetization below 50 K (Fig. 2
and Fig. 3). Since the magnetic order of "super-ferromagnetic" cluster is
local anisotropy effect, this effect is very prominent in low field
magnetization data (Fig. 2) and TRM vs T data (Fig. 8) and in high field
measurement this effect is suppressed by the global effect of small size
clusters (see Fig. 3 and Fig. 8 for 1 Tesla).\\
At this juncture, we would like to add a note of
caution for measurements performed in magnetometers using
superconducting magnets, which have remanent field due to trapped flux when 
manget current is made zero. The remanant field would be large 
(usually about -50 G and could be even higher) in high field (12 Tesla) 
magnetometers such as VSM using Nb$_3$Sn magnets. 
If sufficient caution is not excercised during application of such
magnetometer, one might get negative magnetisation at low field 
values. An example of such an observation encountered by us
is shown in Fig. 9. Therefore, sufficient care should be taken \cite{Lowfield} 
to ensure that the field is truly zero by applying compensating field of exact 
magnitude for the magnetic response in delicate systems such as spin glass.  

\section{Conclusions}
On the basis of dc magnetic measurements of
Co$_{0.2}$Zn$_{0.8}$Fe$_{1.95}$Ho$_{0.05}$O$_{4}$ spinel oxide, we made the
following conclusions:\\
(i) The antiferromagnetic B sublattice superexchange (J$_{BB}$) interactions 
are strongly modified into dominant ferromagnetic interactions due to random 
B site occupancy of Ho$^{3+}$ ions. Consequently, the B site clusters are 
grouped into "super-ferromagnetic" clusters with Ho$^{3+}$ moments and
small clusters without Ho$^{3+}$ moments.\\
(ii) "Super-ferromagnetic" clusters, a fraction in number of total B site
clusters, show superparamagnetic blocking below 180 K along different local 
axes determined by anisotropy contribution of Ho$^{3+}$ and contribute
ferromagnetic order below 50 K due to spin re-orientation effect of Ho$^{3+}$.
On the otherhand, small clusters show spin glass behaviour below
$\approx$ 120 K. In presence of high magnetic field (even at very low field FC
magnetization) the local effect of Ho$^{3+}$ is suppressed by the
field induced ferromagnetic order. As a
result, the sample shows re-entrant magnetic behaviour with
paramagnetic to ferromagnetic state below T$_C$ $\approx$ 225$\pm$ 5 K and
ferromagnetic to cluster spin glass state below T$_f$ $\approx$ 120K.\\
(iii) The random distribution and
competetion between superexchange interactions amongst various magnetic
moments (Fe$^{3+}$, Co$^{2+}$ and Ho$^{3+}$) shows spin glass behaviour 
in the sample, where as the dominent ferromagnetic order inside the
"super-ferromagnetic" clusters and spin canting effect inside the small
clusters show field induced ({\it not infinite long range})
ferromagnetic order in the sample.\\
(iv) 
The existence of cluster size distribution and competetion of different magnetic
order deviate the spin glass dynamics of the sample both from 
typical character of vector mean-field predicted Ising and 
Heisenberg SG. However, the single domain nature, along with anisotropy 
effect, of the "super-ferromagnetic" clusters show the sample to be close to 
Ising SG or randomly anisotropic Heisenberg SG.\\
(v) Finally, the unusual TRM maximum about 170 K is attributed to the 
competetion between ferromagnetic order and superparamagnetic blocking related
to anisotropy effect of Ho$^{3+}$ moments in "super-ferromagnetic" clusters. 
\vspace{0.3 true cm}\\
\noindent Acknowledgement:
RNB thanks the Council of Scientific and Industrial
Research (CSIR, New Delhi, India) for providing fellowship {F.No.
9/489(30)/98-EMR-I].
\newpage
\centerline{Figure Captions}
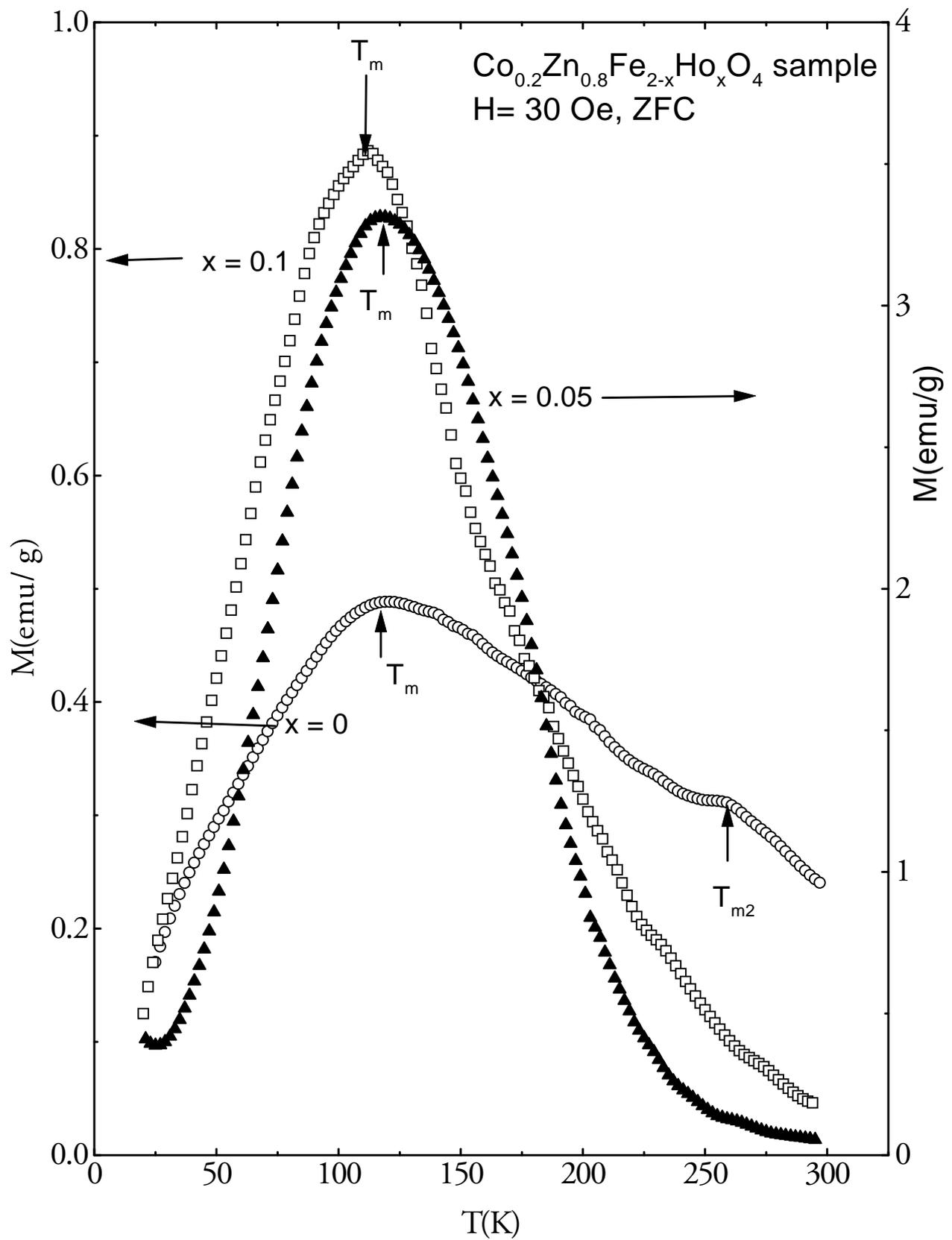
Fig. 1 M vs T data at 30 Oe in ZFC mode for
Co$_{0.2}$Zn$_{0.8}$Fe$_{1.95}$Ho$_{0.05}$O$_4$ sample. The left and right
arrow indicates the magnetization (M) axis.\\
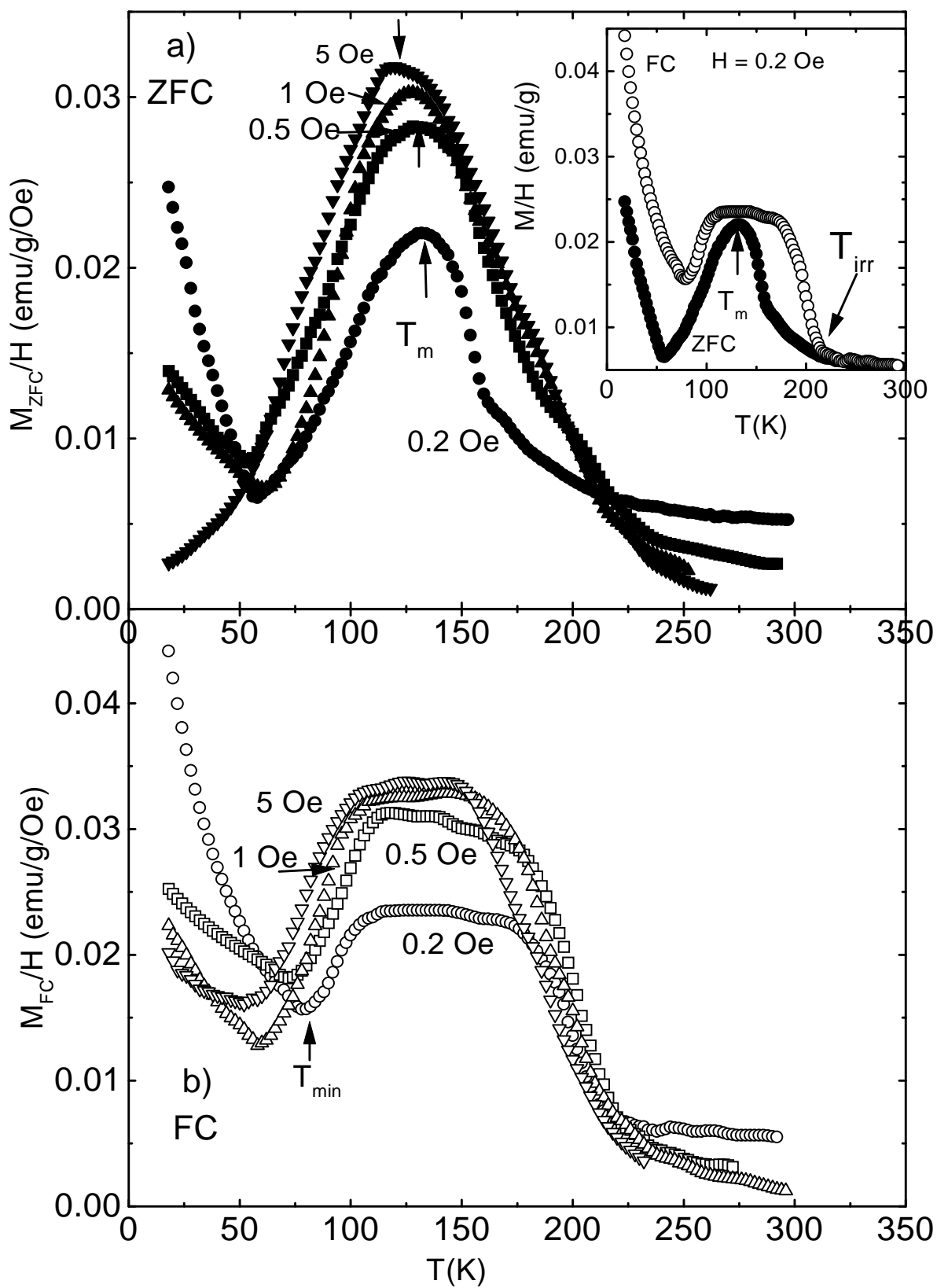
Fig. 2 a) ZFC, b) FC magnetization of
Co$_{0.2}$Zn$_{0.8}$Fe$_{1.95}$Ho$_{0.05}$O$_4$ sample at 0.2 Oe to 5 Oe.
Inset of a) shows the ZFC and FC magnetization at 0.2 Oe.\\
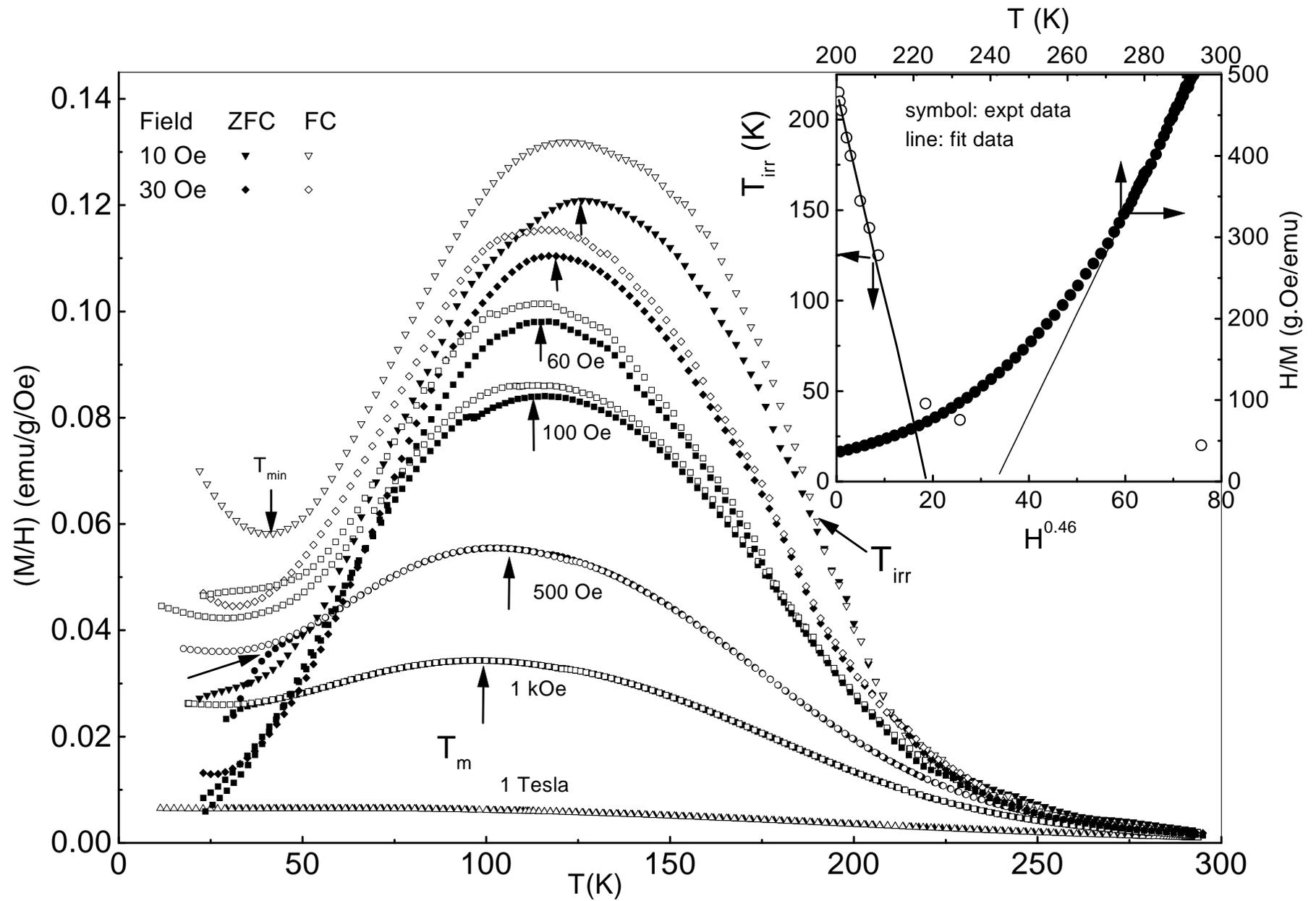
Fig. 3 ZFC and FC magnetization vs T at different fields. T$_m$: MZFC peak
temperature, T$_{irr}$: temperature below which MFC shows irreversibility with
terpect to MZFC, T$_{min}$: temperature where MFC shows minimum. Inset shows 
T$_{irr}$(H) (left scale) and H/M vs T data (right scale).\\ 
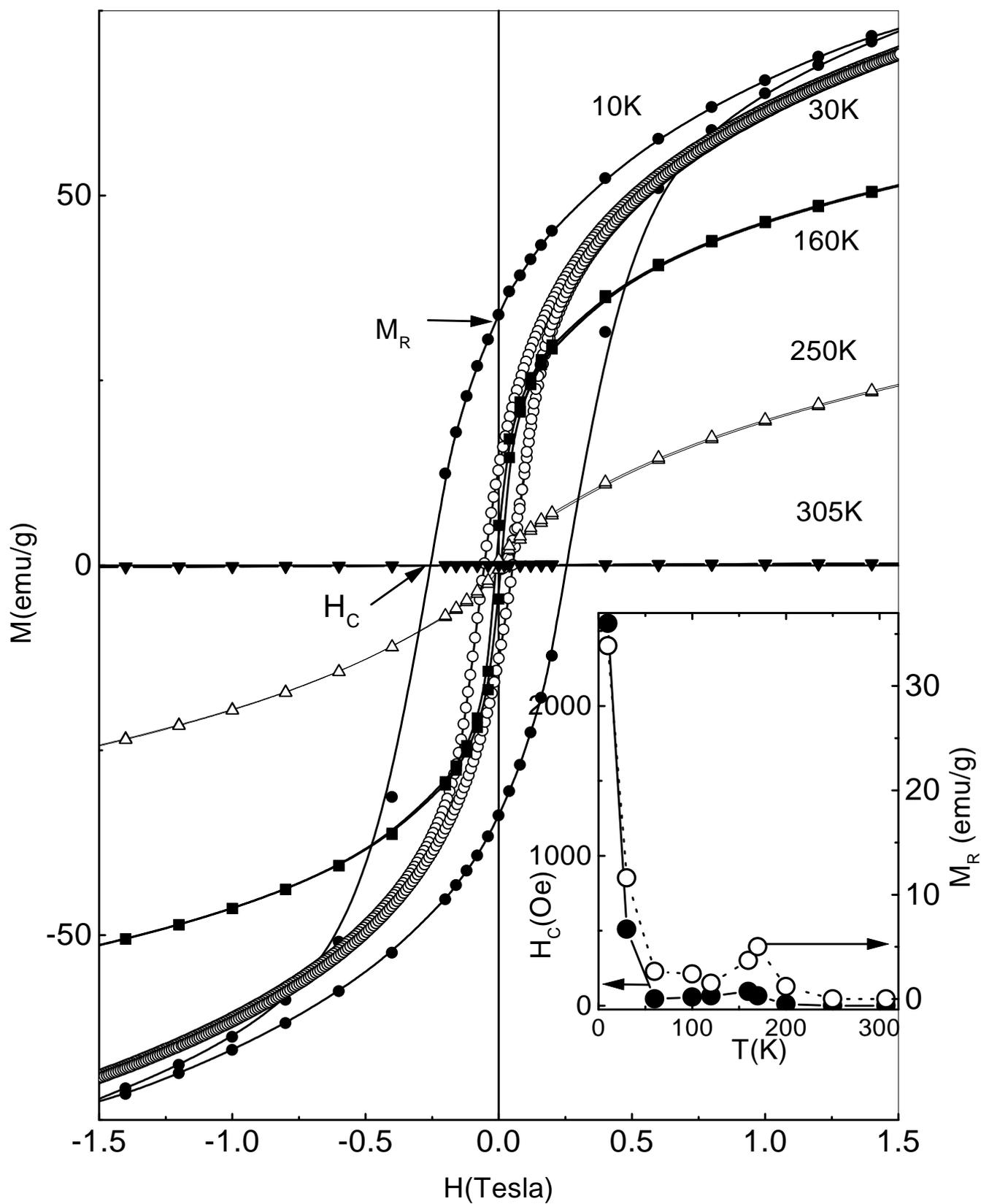
Fig. 4 Inset shows temperature dependence of coercive field H$_C$ (left
scale)
and remanent magnetization M$_R$ (right scale). The main panel shows the
hysteresis data at different temperatures.\\
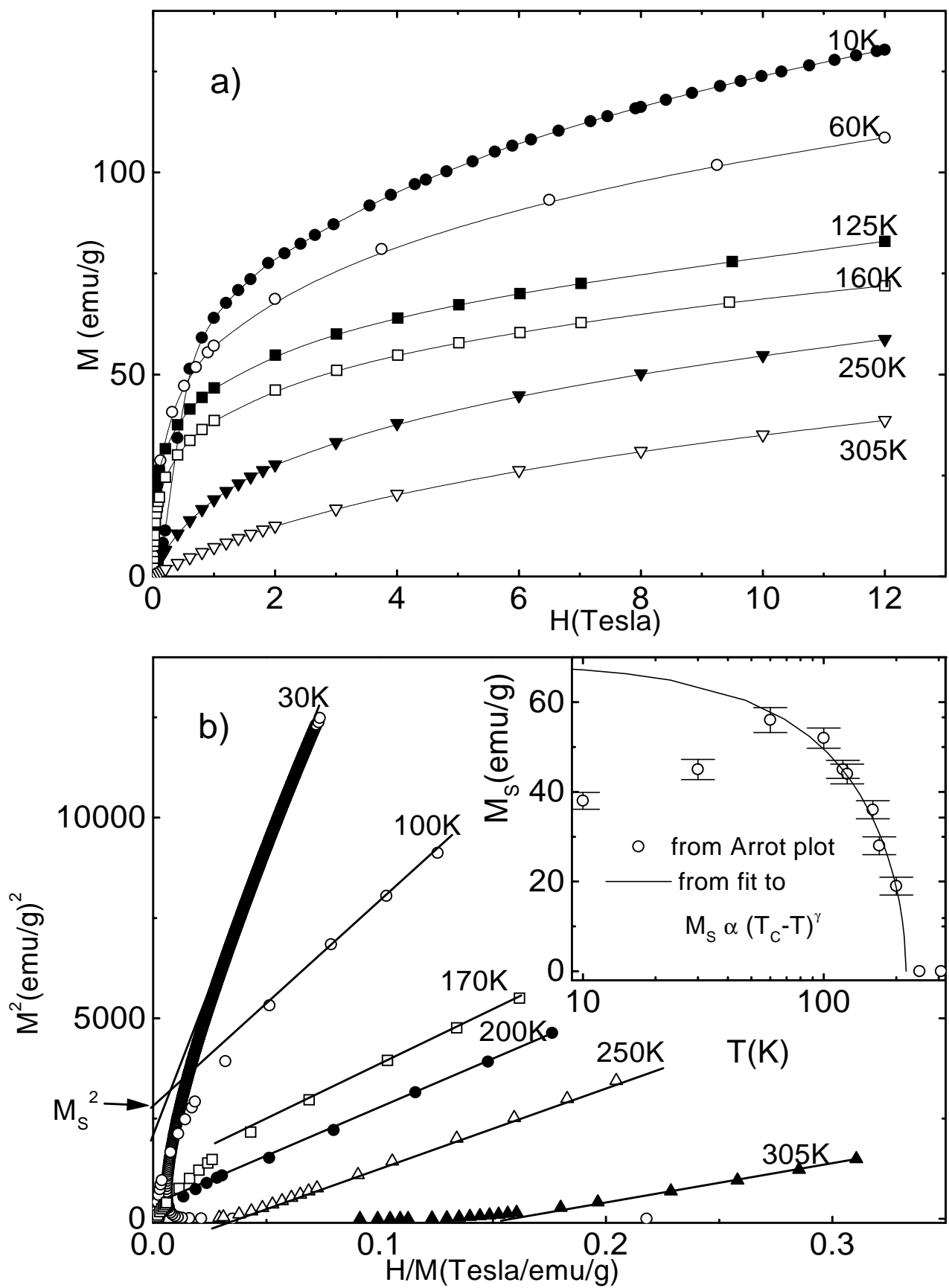
Fig. 5 a) M vs H data at different temperatures. solid lines guide to eye.
b) Arrot plot (M$^2$ vs H/M) at different temperatures. Inset (b) shows the
spontaneous magnetization (M$_S$), obtained from Arrot plot and fit data
(solid line).\\
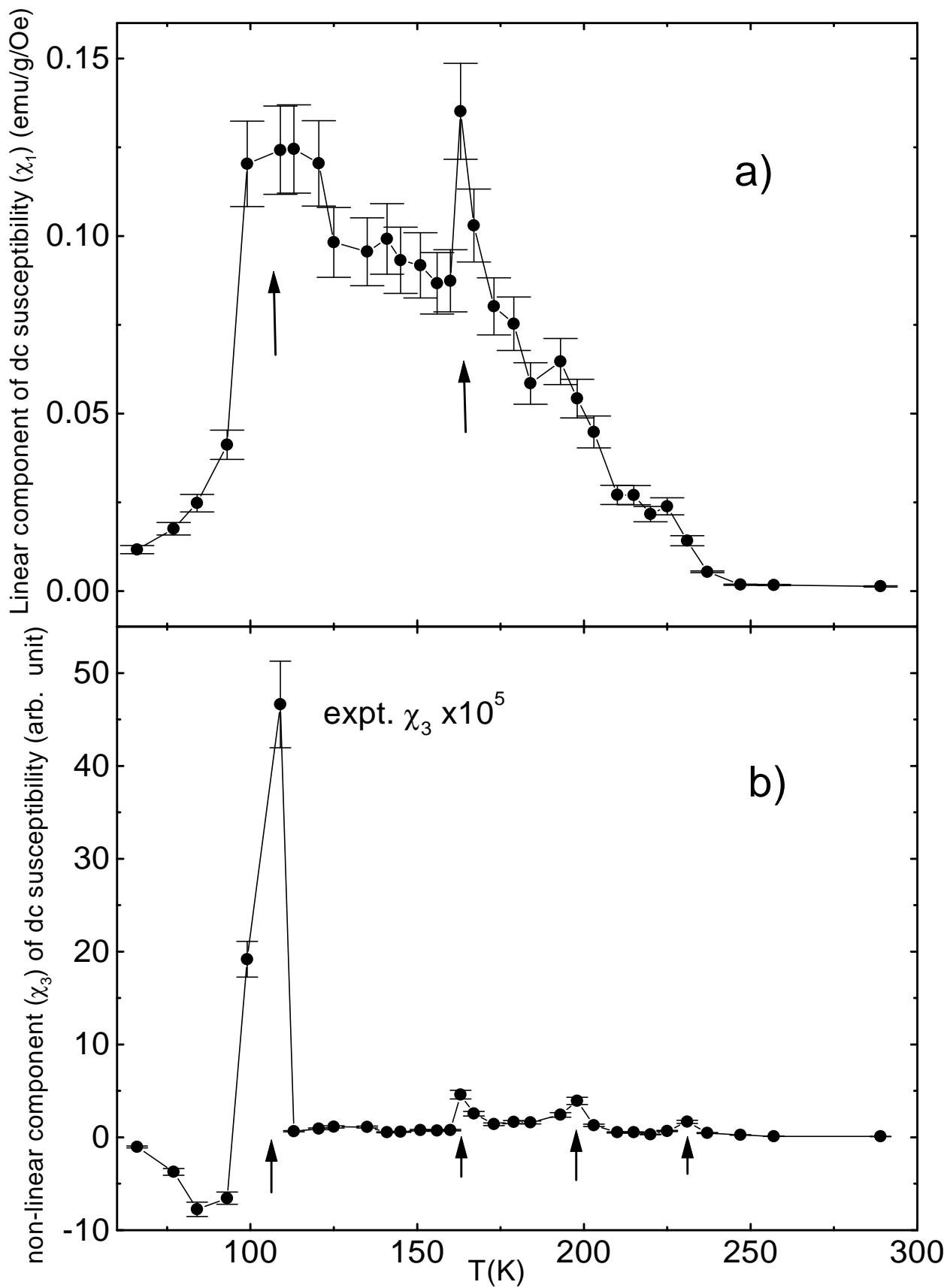
Fig. 6 Temperature dependence of linear and first component of non-linear
dc susceptibility, obtained from M vs H (0 Oe to 100 Oe) data,
for Co$_{0.2}$Zn$_{0.8}$Fe$_{1.95}$Ho$_{0.05}$O$_4$ spinel oxide. Note the
scale factor for $\chi_3$ and the arrows indicate the possible transition
temperatures. data are plotted with 10\% error.\\ 
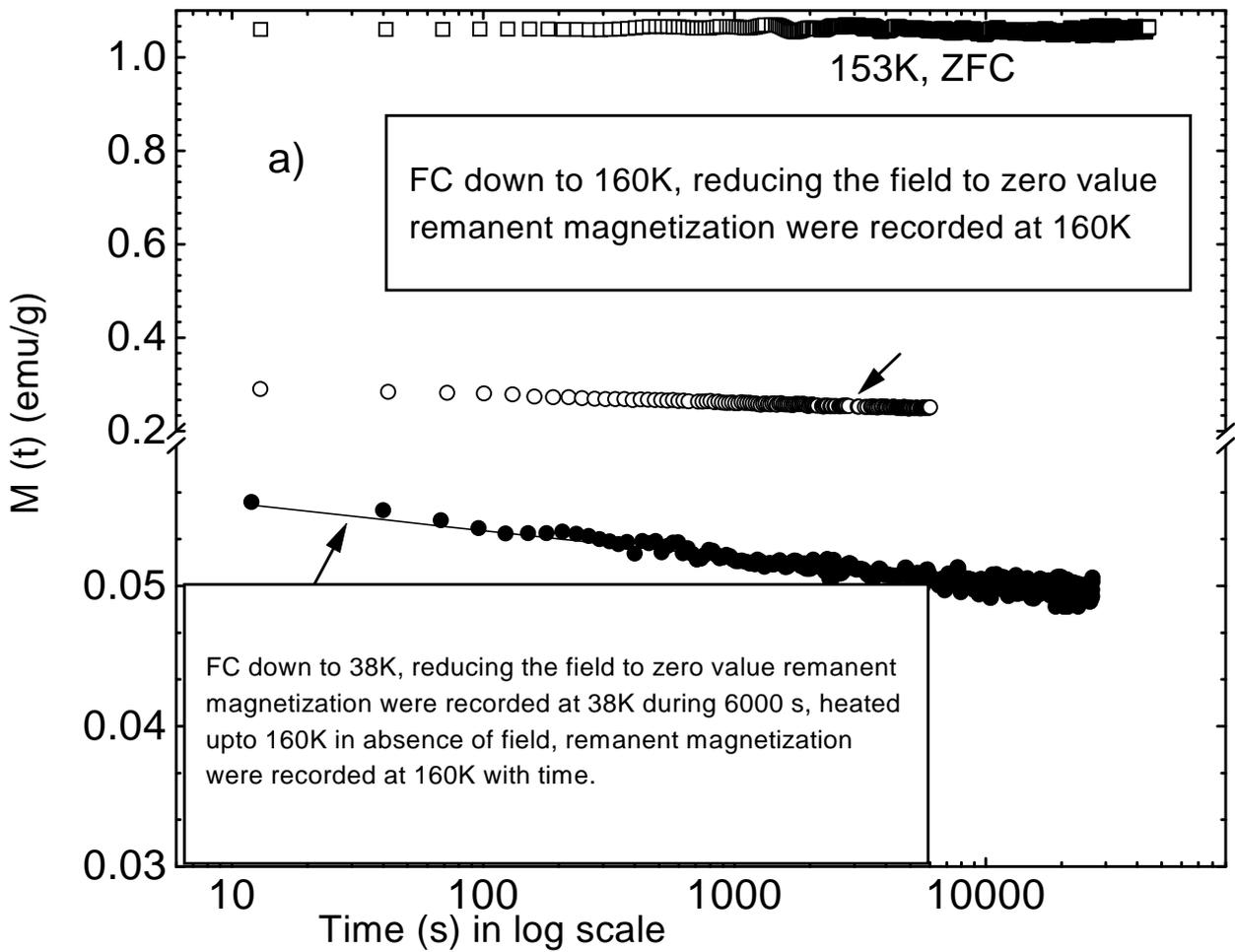
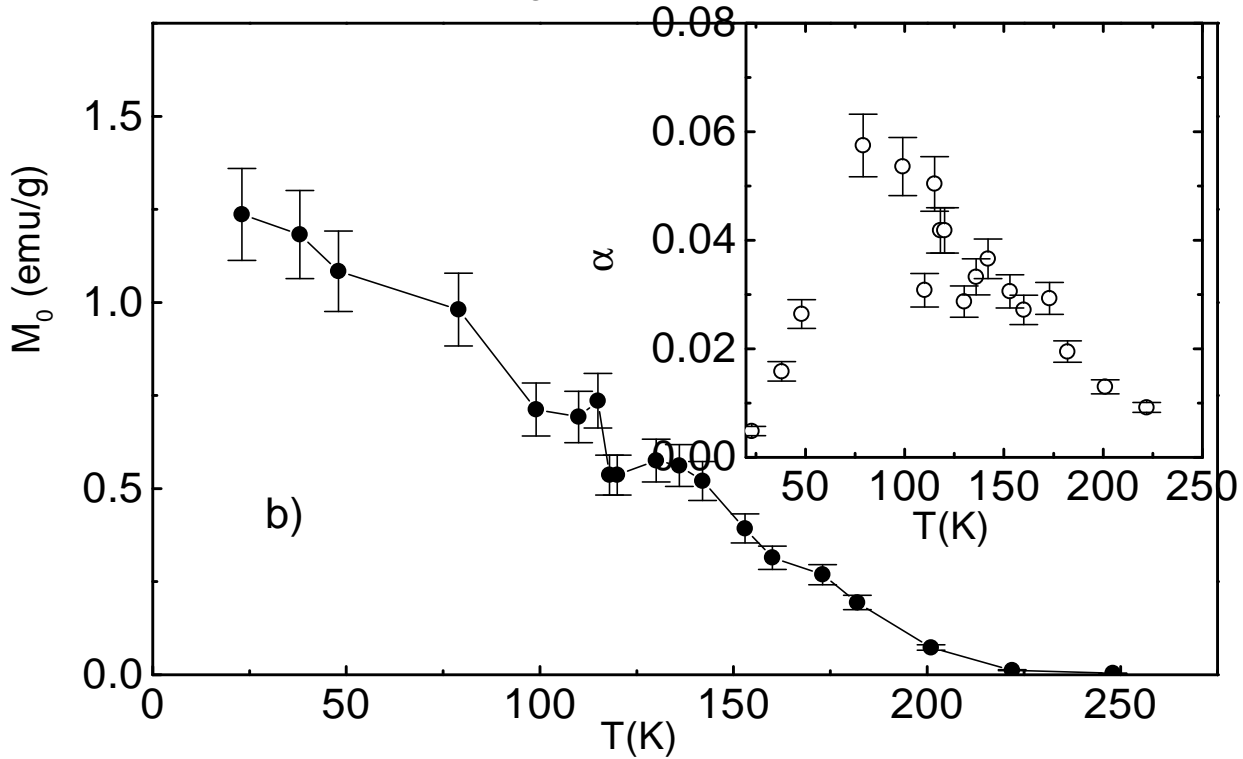
Fig. 7 a) Time dependence of magnetization (ZFC: 153K, FC: 160K). b)
Temperature dependence of the M$_0$ obtained by fitting the relaxation data.
Inset shows the temperature dependence of the $\alpha$ obtained by fitting.\\
Fig. 8 
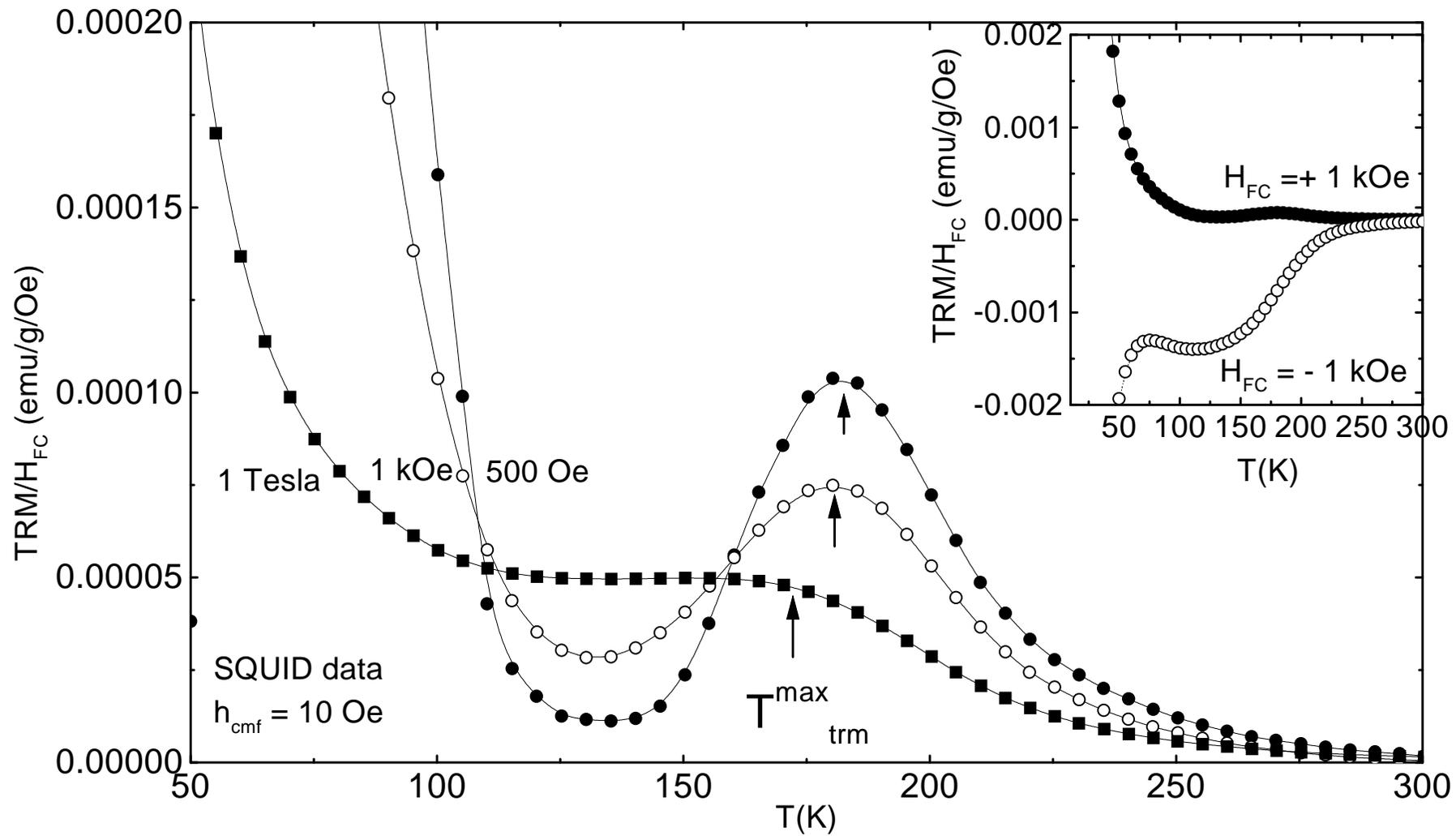
TRM vs T data for 500 Oe to 1 Tesla cooling field. The measurement was 
performed using SQUID with compensating field +10 Oe.
Inset shows TRM vs T data measured at +1 Tesla and -1 Tesla cooling
field.\\ 
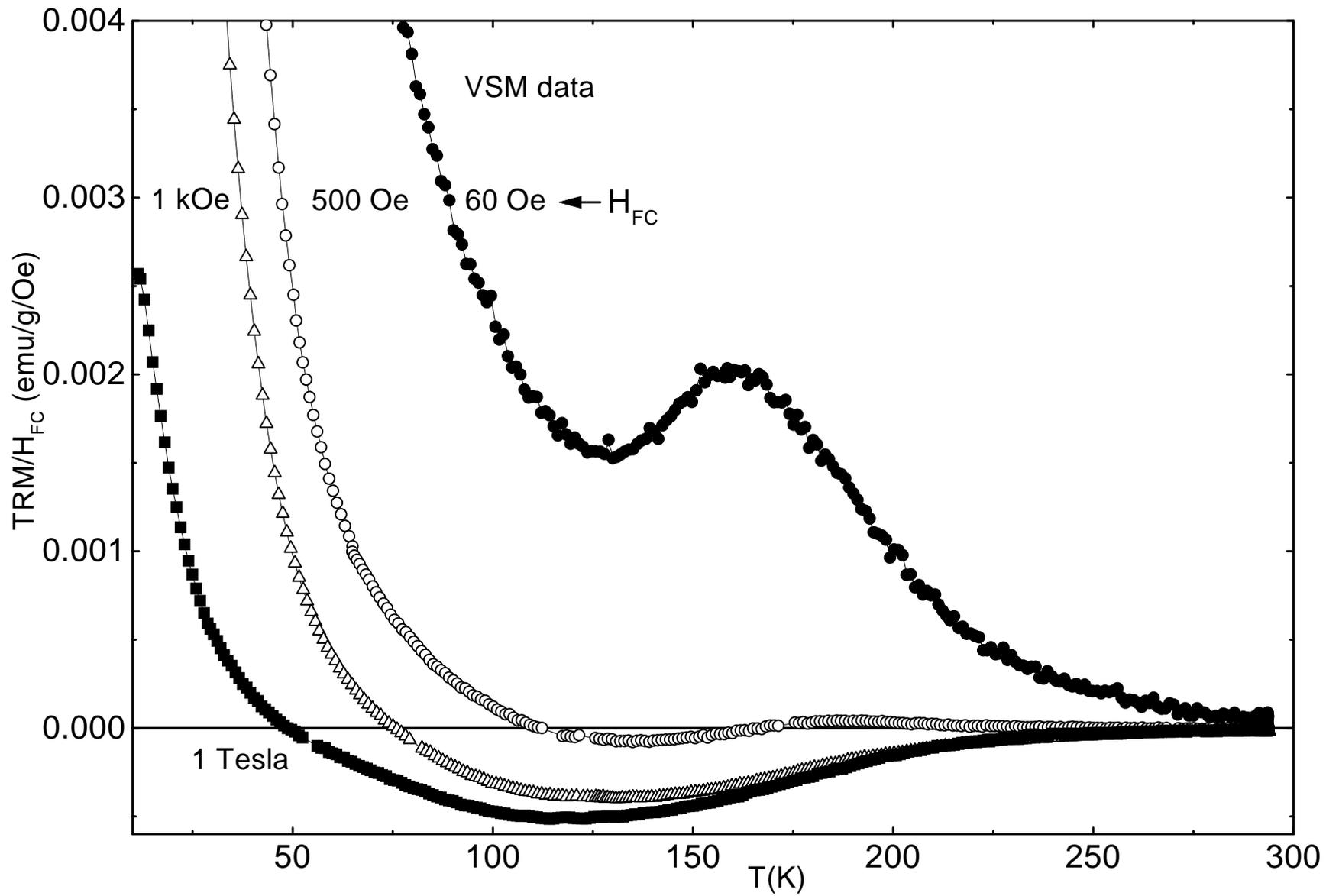
Fig. 9 TRM vs T data measured at different cooling field (H$_{FC}$) using VSM 
magnetometer.


\begin{thebibliography}{99}
\bibitem{Lowfield} see Fig. 2 for example, how a very small field can make a 
large difference to the mangetization. This measurement
performed in a air core copper magnet having no remanent
field.
\end{thebibliography}
\end{document}